\documentclass[
twocolumn,
showpacs,
preprintnumbers,
amsmath,
amssymb,
apl,
]{revtex4-1}

\usepackage{graphicx}
\usepackage{dcolumn}
\usepackage{bm}

\begin{document}

\preprint{APS/123-QED}

\title{
Coherent transport through a double donor system in silicon
}

\author{J. Verduijn$^{1}$}
\email{j.verduijn@tudelft.nl}
\author{G.C. Tettamanzi$^{1}$}
\author{G.P. Lansbergen$^{1}$}
\author{N. Collaert$^{2}$}
\author{S. Biesemans$^{2}$}
\author{S. Rogge$^{1}$}

\affiliation{$^{1}$Kavli Institute of Nanoscience, Delft University of Technology, Lorentzweg 1, 2628 CJÊ Delft, The Netherlands}
\affiliation{$^{2}$InterUniversity Microelectronics Center (IMEC), Kapeldreef 75, 3001 Leuven, Belgium}

\date{\today}

\begin{abstract}
Quantum coherence is of crucial importance for the applicability of donor based quantum computing. In this Letter we describe the observation of the interference of conduction paths induced by two donors in a nano-MOSFET resulting in a Fano resonance. This demonstrates the coherent exchange of electrons between two donors. In addition, the phase difference between the two conduction paths can be tuned by means of a magnetic field, in full analogy to the Aharonov-Bohm effect. 
\end{abstract}

\pacs{nnn}

\maketitle

Dopants gained attention in the past years due to their potential applicability in (quantum) computation architectures using the charge or spin degree of freedom \cite{Kane1998, Hollenberg2004}. In a bulk system, dopants provide long spin-coherence times \cite{Eriksson2004}. Furthermore, the natural potential landscape of a dopant is very robust and exactly reproducible. However, for practical applications, the dopants need to be embedded in nanostructures allowing manipulation and readout of the (quantum mechanical) state \cite{Kane1998, Hollenberg2004}. This modifies their bulk properties significantly \cite{Sellier2006, Lansbergen2008, Rahman2009} and thus requires new experiments to probe quantum coherent electron exchange and electronic properties such as the level spectrum. In this Letter, we study transport signatures that provide information about the electronic coherence. In particular, we report the observation of phase coherent transport of electrons through two physically separated donors, resulting in Fano resonances at low temperature.
\par
Our devices are silicon FinFETs with a boron-doped channel and a poly-silicon gate wrapped around the channel \cite{Sellier2007}. Few arsenic dopants (n-type) diffuse into the p-type channel from the highly doped source/drain regions and modify the transport characteristics \cite{Sellier2006}. Recently, it has been shown that the level spectrum of isolated dopants can be determined by means of low temperature transport spectroscopy \cite{Lansbergen2008}, but since this work relies on statistics to find a single dopant in the transport, there are also devices that exhibit multi-dopant transport. In fact, transport occurs through a single dopant only in about 1 out of 7 devices with a fixed gate length and channel height of 60 nm and channel widths varying between 35 nm and 1 $\mu$m \cite{Lansbergen2008}. All other devices show multi-dopant transport or no signatures of dopants at all \cite{Sellier2006}. The device we discuss in this Letter has a gate length of 60 nm and channel width of 35 nm.
\par
\begin{figure}[!ht]
\begin{center}
\includegraphics[width=0.4\textwidth]{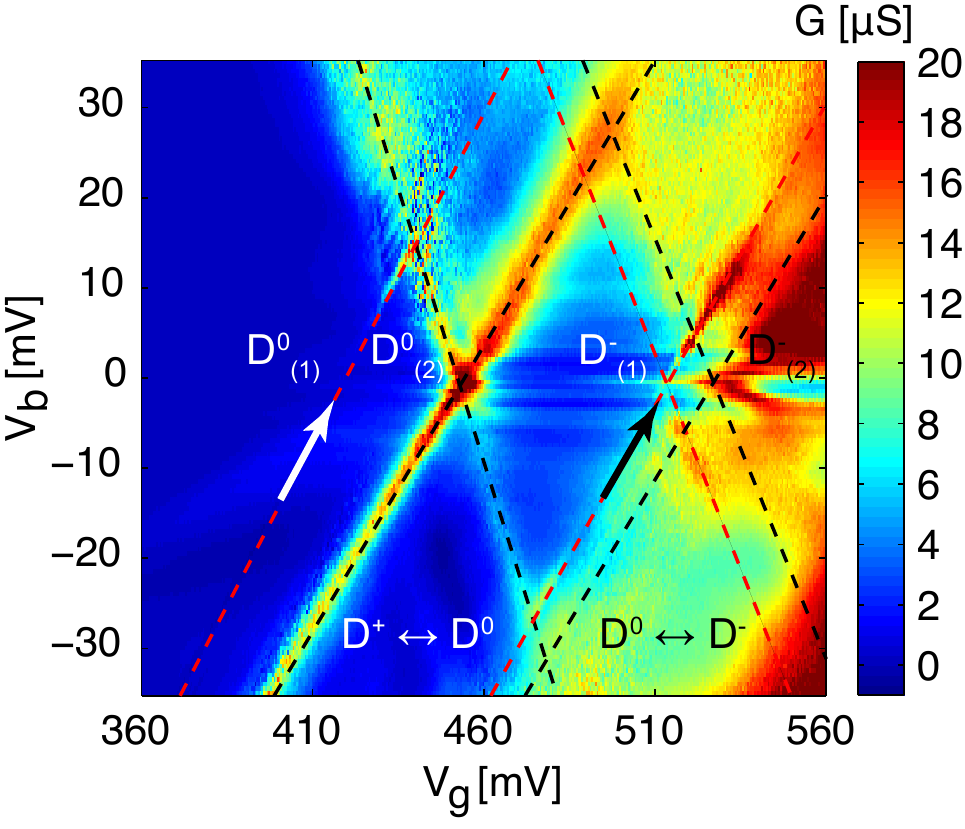} \caption{The drain current and the differential conductance are measured in a three-terminal configuration. These data are plotted in a differential conductance stability diagram and reveal that the transport at 0.3 K is largly determined by a single As donor. Regions where transport occurs through the neutral $D^0$ state and through the $D^-$ state can be distinguished (indicated by the black dashed lines). In addition, we observe a two narrow Fano lines in the vinicity of the $D_{(2)}^0$ and $D_{(2)}^-$  transport features (indicated by arrows). Inside the Coulomb diamond there is a zero-bias feature visible which can be attributed to a Kondo effect.}
\label{fig:f1}
\end{center}
\end{figure}
We measure the dc characteristics of our devices, namely the drain current, $I$, and the differential conductance, $G=\text{d}I/\text{d}V_b$, versus the gate voltage, $V_g$, and bias voltage, $V_b$, in a three-terminal configuration. The differential conductance is measured using a lock-in technique with a 50 $\mu$V sinusoidal ac excitation at 89 Hz, superimposed on the dc bias component. These obtained data can be plot in a stability diagram, a two-dimensional color-scale plot with the gate voltage and bias voltage on the axes. From the stability diagram, measured at low temperature ($\lesssim4.2$ K), one can typically extract information such as the level spectrum of the donor and the energy needed to add a second electron to the system, the charging energy \cite{Lansbergen2008}.
\par
\begin{figure}[!ht]
\begin{center}
\includegraphics[width=0.4\textwidth]{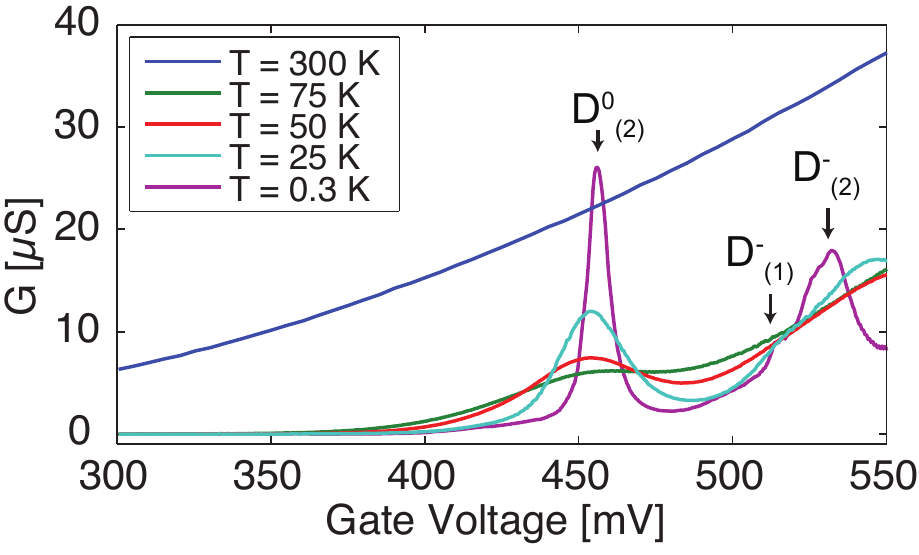} \caption{A trace at zero bias in gate voltage versus temperature is taken. We see a strong increase of the height of the Coulomb oscillations with decreasing temperature. This proves there is coherence in the transport of electrons \cite{Beenakker1991}.}
\label{fig:f2}
\end{center}
\end{figure}

Figure \ref{fig:f1} shows the differential conductance as a function of $V_b$ and $V_g$. We observe two triangular regions with a non-zero differential conductance due to direct tunneling processes through donor states in the FinFET channel. The corresponding resonances at $V_b=0$ mV are denoted $D^0_{(2)}$ and $D^-_{(2)}$ in Fig. \ref{fig:f1} and Fig. \ref{fig:f2}. At lower gate voltage ($V_g\sim$455 mV), direct transport regions, bound by black dashed lines, can be distinguished. Here the donor is alternating between the ionized ($D^+$) and neutral state ($D^0$) while electrons traverse the donor one-by-one. At higher gate voltage ($V_g\sim$530 mV) a second region is visible where the donor alternates between the $D^0$ and the negatively charged state ($D^-$). The diamond shaped area in between marks Coulomb blockade of the donor with a fixed number of electrons. To investigate the mode of transport we show conductance traces in $V_g$ at zero $V_b$ as a function of temperature (Fig. \ref{fig:f2}). Lowering the temperature from 75 K to 50 K already results in an increase of the resonance denoted by $D^0_{(2)}$, indicating no internal relaxation occurs \cite{Beenakker1991}. At base temperature (0.3 K) the maximum conductance even exceeds the room temperature value, approaching $0.67e^2/h$. These observations indicate that there is phase coherence at low temperature \cite{Beenakker1991}. Considering the addition energy of about 35 meV, the presence of a zero bias Kondo line in the Coulomb blockade region and the Zeeman shift of the $D^0_{(2)}$ resonance, these features are most likely to be due to an arsenic donor close to the Si/SiO$_2$ channel interface \cite{Sellier2006, Lansbergen2008, Sellier2007, Lansbergen2009}.
\par
\par
\begin{figure}[!ht]
\begin{center}
\includegraphics[width=0.4\textwidth]{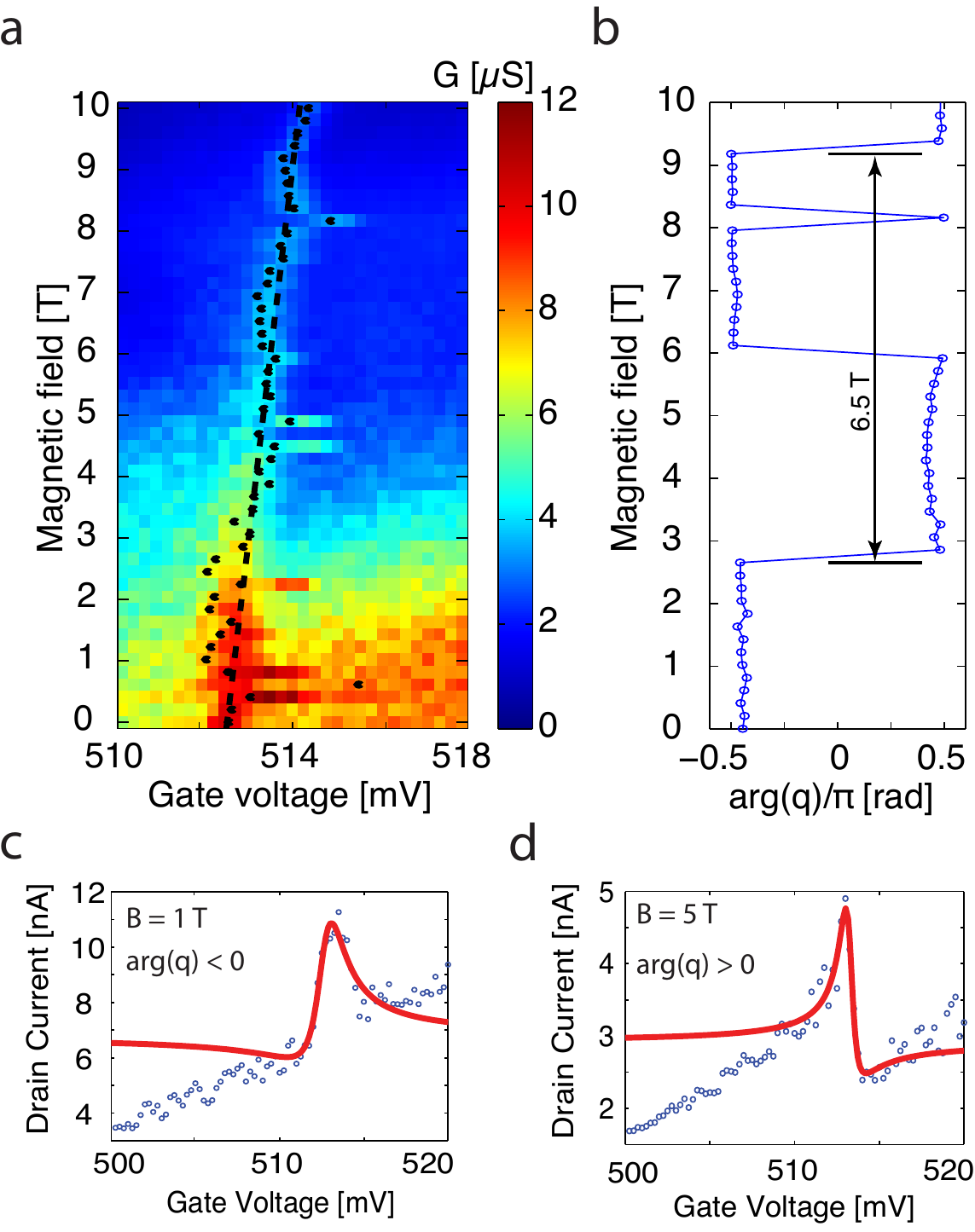} \caption{(a) Sweeping the gate voltage at different magnetic fields reveals, by the shift to higher gate voltages, that the Fano resonance carries spin down electrons which is consistent with the state being a charged donor state $D^-$ \cite{Sellier2006}. We fit these traces using a fenemenological formula and obtained the complex Fano parameter $q$ (see main text). Furthermore, we fit a linear function to the peak positions (black dots) and convert this to energy, using the gate coupling $\alpha$, to find the shift of the peak as a function of the field, we find $0.12\pm0.02$ meV/T consistent with a shift dominated by the Zeeman energy \cite{Weis1993}. (b) We plot the argument of $q$ ($\arg(q)$) to quantify the magnetic field dependence, in particular, the symmetry transition of the peak. The period of this symmetry transition is found to be 6.5 T. (c, d) The Fano formula fits well, $R^2\sim0.9$, and the peak shows a symmetry transition as a function of the magnetic field.}
\label{fig:f3}
\end{center}
\end{figure}

In addition to the clearly visible $D^0_{(2)}$ and $D^-_{(2)}$ features discussed above, two much fainter resonances are visible, denoted by the arrows in Fig. \ref{fig:f1} and labeled as $D^0_{(1)}$/$D^-_{(1)}$. At high bias, $V_b>0$, these resonances develop in faintly visible triangular regions, due to first order sequential tunneling (red dashed lines). This provides a way to extract the charging energy of the localized state at the origin of these resonances, we find $\sim$35 meV. Furthermore, the resonance at $V_g\sim510$ mV ($D^-_{(1)}$) shows a linear shift towards higher gate voltages of about $0.12\pm0.02$ meV/T when magnetic field, $B$, between 0 T and 10 T in the direction of the channel is applied (Fig. \ref{fig:f3}a). This is the Zeeman shift expected for a spin singlet state \cite{Weis1993, Sellier2006}. Altogether this makes us confident that the origin of the resonances is a second donor. It must be noted that the $D^0_{(1)}$ resonance is too weak compared to the background to observe any shift under magnetic field unambiguously.
\par
A trace in $V_g$ around the $D^-_{(1)}$-resonance (Fig. \ref{fig:f3}c and \ref{fig:f3}d) reveals that this resonance has a Fano line shape \cite{Fano1935}. We suggest that the two conduction paths, induced by the donors 1 and 2, add in a coherent way, resulting in destructive or constructive interference, and in this way give rise to a Fano resonance. Fano resonances have been observed in a wide range of physical systems \cite{Miroshnichenko2009}. To observe this effect, a path with a constant or slowly varying phase that interferes with a path with a rapid phase variation is required. In our system the phase varies as a function of the energy difference between the donor state and the chemical potential of the contacts \cite{Gores2000}. The gate allows us to tune the energy of the localized states and thereby, effectively, the transport phase. The remainder of this Letter discusses the nature of the interference in the device.
\par
To gain more insight we measure differential conductance traces in $V_g$ around the $D^-_{(2)}$ Fano resonance while applying a magnetic field, $B$, parallel to the FinFET channel between 0 and 10 T (Fig. \ref{fig:f3}a). We observe that the line shape changes as we increase the field and even changes symmetry in an alternating fashion (Fig. \ref{fig:f3}c and \ref{fig:f3}d). This indicates that we tune the phase difference between (at least) two coherent current paths \cite{Yacoby1995}. For this to occur, the paths must be physically separated and form a closed loop in such a way that a net magnetic flux can pierce the formed loop and modify the phase difference by the Aharonov-Bohm (AB) effect \cite{Aharonov1959}. Therefore, we conclude that we probe two physically separated donors in a phase coherent way.
\par
In order to make this effect more quantitative, we fit the traces in gate voltage taken at magnetic fields between 0 T and 10 T to a phenomenological formula \cite{Fano1935} \[G\left(\epsilon\right)=G_F\frac{{\left | \epsilon+q\Gamma/2\right |}^2}{\epsilon^2+(\Gamma/2)^2}.\] Where $\epsilon$, $\Gamma$ and $G_{F}$ are the detuning of the resonance, tunnel coupling and a pre-factor respectively. The detuning can be related to the gate voltage via the gate coupling $\alpha$, defined as $\epsilon=\alpha (V_g-V_{g,0})$, where $V_{g,0}$ is the position of the resonance. The gate coupling, $\alpha$, can be obtained from the stability diagram by dividing half the height of the Coulomb diamond by its width \cite{Sellier2007}. We take the Fano parameter $q=q_x+iq_y$ as a complex number to account for the non-coherent contribution to the conductance \cite{Clerk2001}. The argument of the Fano parameter, $\arg{(q)}=\arctan{(q_y/q_x)}$, varies between $\pm\pi/2$ as a function of the magnetic field as can be seen in Fig \ref{fig:f3}b. This reflects the symmetry change of the resonance. Since the symmetry change in the resonance is periodic in the flux quantum $\Phi_0=h/e$, by the nature of the AB effect \cite{Yacoby1995}, we can determine the projected surface area of the loop formed by the two current paths. Using a period of 6.5 T from the data (Fig. \ref{fig:f3}b) we find a surface area $A\sim6.3\cdot10^{-16}$ m$^2$. This corresponds to a circular loop with a diameter of $\sim$28 nm, which is a realistic size considering the dimensions of our structure. Also the stability diagram (Fig. \ref{fig:f1}) shows that there is no direct coupling between the donor, since this would result in hybridization of the orbitals of both donors, reflected by a shift in the stability diagram. Therefore we conclude that the inter-donor distance must be $\gtrsim$ 20 nm \cite{Koiller2006}. Supported by the found projected loop size, we argue that this is also consistent with the coherent transfer of electron between two independent donors.
\par
Furthermore, we observe that the background as well as the Fano resonance decreases with magnetic field (Fig. \ref{fig:f3}b). Well away from the resonance, e.g. at $V_g\sim475$mV), the background is due to the Kondo effect, and is thus quenched by the magnetic field \cite{Meir1993}. Therefore, we speculate that the Fano resonance is the results of interference between a Kondo transport channel and a direct transport processes.
\par
In summary, we demonstrate phase coherent exchange of electrons between two donors at low temperature. This is a key ingredient to single-donor quantum device applications. The observation of a Fano resonance, due to the interference between two conduction paths induced by the two donors, is a proof of phase coherence in our device. We speculate that this interference effect originates from the interplay between a Kondo- and a direct transport channel. The phase difference between the two conduction paths can be tuned by means of a magnetic field, analogues to the AB-effect. This analysis indicates that the distance between these donors is on the order of the device dimensions. Consistent with this, the transport measurements show no signs of direct interaction between the two donors. Thus, we conclude that the donors are physically separated and only coherently coupled in transport.

\bibliography{DonorFano_final}

\end{document}